\documentclass[epj]{svjour}
\usepackage{graphics}
\begin{document}
\title{Crossover phenomenon in the performance of an Internet search engine}
\author{Lucas Lacasa$^1$, Jacopo Tagliabue$^2$, and Andrew Berdahl$^{3}$}% etc
\institute{Departamento de Matem\'{a}tica Aplicada y Estad\'{i}stica, ETSI Aeron\'{a}uticos, Universidad Polit\'{e}cnica de Madrid, \and Department of Philosophy, Universit\`{a}
Vita-Salute San Raffaele, Milan, Italy \and Department of
Ecology and Evolutionary Biology, Princeton University, United
States}

\date{\textbf{Working paper Santa Fe Institute CSSS09}}
\abstract{In this work we explore the ability of the Google search
engine to find results for random $N-$letter strings. These random
strings, dense over the set of possible $N-$letter words, address
the existence of typos, acronyms, and other words without semantic
meaning. Interestingly, we find that the probability of finding
such strings sharply drops from one to zero at $N_c=6$. The
behavior of such order parameter suggests the presence of a
transition-like phenomenon in the geometry of the search space.
Furthermore, we define a susceptibility-like parameter which
reaches a maximum in the neighborhood, suggesting the
presence of criticality. We finally speculate on the possible
connections to Ramsey theory.}

%
%\PACS{
 %     {PACS-key}{discribing text of that key}   \and
 %     {PACS-key}{discribing text of that key}
  %   } % end of PACS codes
 %end of abstract
%
\titlerunning{Crossover in search engine performance}
\maketitle
\section{Introduction}
Computer science and physics, although different disciplines in
essence, have been closely linked since the birth of the former. More recently,
 computer science has met together with statistical
physics in the so called combinatorial problems and their relation
to phase transitions and computational complexity (see \cite{SFI}
for a compendium of recent works). More accurately, algorithmic
phase transitions (threshold property in the computer science
language),
 i.e. sharp changes in the behavior of some
computer algorithms, have attracted the attention of both
communities \cite{FIRST,MERT,SAT,TF,TF2,TF3,TF4,TF5}. It has been
shown that phase transitions play an important role in the
resource growing classification of random combinatorial problems
\cite{TF}. The computational complexity theory is therefore
nowadays experiencing widespread growth, melting different
ideas and approaches coming from theoretical computation,
discrete mathematics, and physics. For instance, there exist
striking similarities between optimization problems and the study
of the ground
states of disordered models \cite{PAR}. \\
Problems related to random combinatorics appear typically in
discrete mathematics (graph theory), computer science (search
algorithms) or physics (disordered systems). The concept of sudden
change in the behavior of some variables of the system is
intimately linked to this hallmark. For instance, Erd$\ddot{o}$s
and Renyi, in their pioneering work on graph theory \cite{erdos},
found the existence of \emph{zero-one} laws in their study of
cluster generation. These laws have a clear interpretation in
terms of phase transitions, which appear extensively in many
physical systems. More recently, computer science community has
detected this behavior in the context of algorithmic problems. The
so called \emph{threshold phenomenon} \cite{SFI} distinguishes
zones in the phase space of an algorithm where the problem is,
computationally speaking, either tractable or intractable. It is
straightforward that these three phenomena can be understood as a
unique concept, in such a way that building bridges between each
other is an
appealing idea.\\
In this work we address the performance of Google's search engine
from a similar point of view. The webpages, blogs, and other text repositories
that compose the Internet contain a huge amount of information, which
is typically encoded in texts -i.e. words with semantic meaning-
of several languages. These words are N-letter strings, where $N$
is not expected to be too large, according to the dictionary.
Eventually, we will find in these information repositories some
words that are not defined in any dictionary. These words can be
typos (typographic errors), acronyms, invented words, etc, that we
will call typos from now on. Since there are many independent
reasons justifying the presence of such typos, as a first
approximation we can suppose that they are the result of a random
process where in every new webpage or blog, with a small
probability a new typo is introduced. The total amount of these
outliers would be, in this case, directly related to the size of
the total text reservoir: Internet should be 'large enough' to
have these structures by pure chance. Now, which is the amount of
these typos as a function of the typo's size? Is there any
characteristic scale for these structures? How can we estimate
such amount? Of course, for every fixed N the are many more words
without a semantic meaning: if we generate a random N-letter
string, with very large probability, this one will not be a real
word, but some kind of typo. Consequently, in order to explore the
presence of typos in Internet, we only need to make queries of
random N-letter strings. Now, are the typos equally distributed as
a function of the typo's size? If these typos are reminiscent of
the real words (for instance, if a typo is just the result of a
word with a permutation/deletion/modification of letters), we
should expect that the presence of N-letter typos is a smoothly
decreasing function of $N$. Will we find such smooth behavior in
this case? In what follows we will present some results suggesting
that the presence of typos is related to a percolation-like
phenomenon, where the probability of finding an N-letter typo
sharply drops from one to zero at a critical value $N_c$. This
latter value is related to the reservoir's size. We finally
speculate on the relation to Ramsey theory, which addresses the
presence of spurious order in random structures.

\section{Automatic random queries} 
\begin{figure}
\resizebox{0.85\columnwidth}{!}{%
 \includegraphics{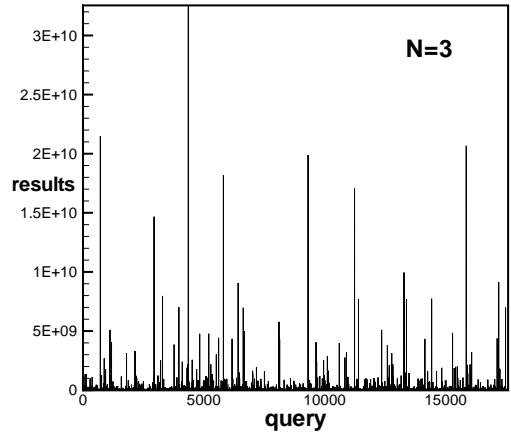}}
\caption{Example of automatic query results for a string of $N=3$
letters. In each query, a $3-$letter string is generated at
random.}\label{N3}
\end{figure}

\begin{figure}
\resizebox{0.85\columnwidth}{!}{%
 \includegraphics{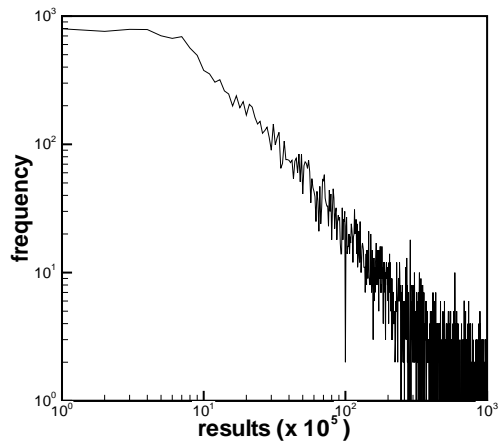}}
\caption{Histogram in log-log of figure \ref{N3}, that plots the
frequency of results. The distribution approximates a uniform one
for small results, characteristic of a random process. The tail
follows a power law: this is reminiscent of the word use
distribution in languages.}\label{N3histo}
\end{figure}

We have done automatic
generated queries to the popular Internet search engine Google.
Fixed a size $N$, we have generated $2\cdot 10^4$ random strings
of $N$ letters and have made the associated queries. Each query
has an associated output, the amount of results $E$. In figure
\ref{N3} we show an example a string of $N=3$ letters. In each
query, a $3-$letter string is generated at random, and we plot E
as a function of the query. In figure \ref{N3histo} we plot, in
log-log, the histogram of such experiment, plotting the frequency
distribution of $E$. The distribution approximates a uniform one
for small results, characteristic of a random process. The tail
follows a power law. If we assume that the presence of typos is
correlated in some way to the presence of real words, we can
deduce that this power law is reminiscent of the word use
distribution in languages, which actually follows a power law in
the statistics of word use in books.


\section{Evidence of critical behavior} 
\begin{figure}
\resizebox{0.85\columnwidth}{!}{%
 \includegraphics{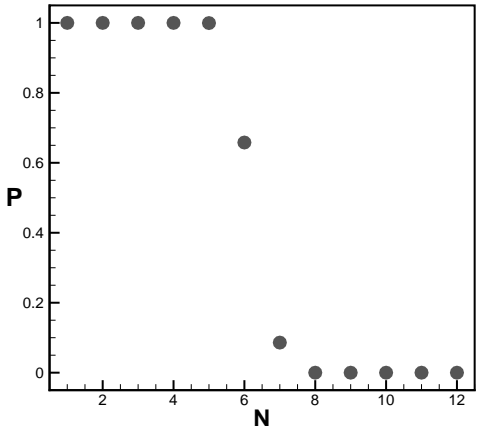}}
\caption{Probability versus N}
\label{P}       % Give a unique label
\end{figure}

We have defined the order
parameter $P$ as the probability of finding a non-null amount of
results whenever making a random N-letter string query to Google.
In practice, and following the definition of $P$ in percolation
theory, in each query we have summed 1 whenever the query shows
non-null results and 0 otherwise, and have finally normalized $P$
over the total number of queries. In figure \ref{P} we have
plotted the values of $P$ versus the number of letters in a
string, $N$, which acts as a control parameter. Below a certain
value $N_c$, the probability of finding a non-null amount of
results is $1$, while above $N_c$, this probability sharply drops
to a value which is very close to zero. Following a geometrical
image, we can understand this behavior as a percolation process in
the space of all possible combinations of n-letter words: while
for $N<N_c$ the majority of these possible combinations are
actually present in the Internet reservoir, and thus we are in the
'percolant phase' where every initial condition (random
combination of $N$ letters) can be found across a non-null amount
of paths, for $N>N_c$ the number of such paths drops to zero. We
expect that this behavior is even more acute for larger sizes of
the Internet reservoir.

\subsection{Susceptibility-like parameter}
\begin{figure}
\resizebox{0.75\columnwidth}{!}{%
 \includegraphics{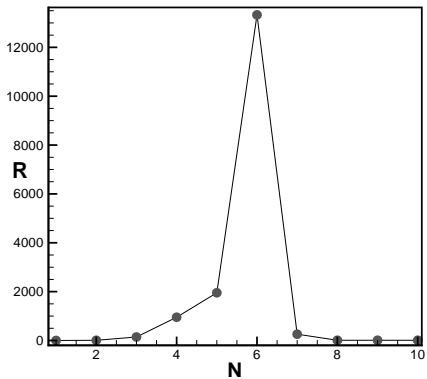}
}
%\vspace{5cm}       % Give the correct figure height in cm
\caption{R versus N}
\label{R}       % Give a unique label
\end{figure}

In order to cast light on the nature of such apparently abrupt
behavior, we need to define the thermodynamically conjugated
variable of the order parameter, that is, a susceptibility-like
parameter that measures the fluctuations of $P$. The so called
canonical measure of self-averaging performs $R$ this task, since
it is defined as the variance of $E$, properly normalized:
$$R=\frac{<E^2>-<E>^2}{<E>^2}$$ As it can be seen in figure \ref{R}, $R$ evidences a
peaked maximum in the neighborhood of the transition point, much
in the vein of a critical transition. This suggests the presence
of criticality in this system.

\section{Possible connections with Ramsey theory}
In a nutshell, Ramsey theory addresses the presence of spurious
order in disordered media. The cornerstone of such theory is the
following: in a set of $M$ elements where no relation of order has
been defined (that is, assuming no correlations between the $M$
elements), one can find with probability 1 hints of order (i.e.
patterns) of arbitrarily size as long as $M$ is large enough.\\
Ramsey theory is, for instance, the reason why we can find several
stars in the sky forming a straight line: this pattern may suggest
the presence of a hidden order, such an extraterrestrial skyway.
However, the fact that there are so many stars in the sky is
sufficient for these kind of geometric patterns to emerge, just by
pure chance.\\
More technically, Ramsey theory addresses the presence of such
patterns in graphs. Concretely, the Ramsey number $r(m,n)$ of a
graph is the minimal number of nodes that a random graph needs to
have in order to contain a clique of order $n$. In our work, a
handwaving analogy could be made: suppose that the Internet
reservoir is the set of $M$ elements. Why do we find random
N-letter strings, which are not obviously -in most of the cases-
true words? There are many reasons: the presence of typos,
acronyms, and other sources of 'randomness'. Now, if for $N<N_c$
the probability of finding such random strings is 1, this suggests
the presence of some order (for instance, as long as the entropy
is low). One could assert that the only reason for this
probability to be 1 is that the Internet is so large (webpages,
blogs, etc) that one can find spurious order, just as in Ramsey
theory. And that given the number of such webpages and blogs, this
spurious order grows until $N_c$. This should be investigated in
depth in future work.

\section{Concluding remarks}
In this work we have shown that the probability of finding a
random N-letter string in Internet shows an abrupt behavior, this
probability being $P\simeq 1$ for $N<6$ while $P\simeq 0$ for
$N>6$. We have interpreted such crossover as a percolation-like
process in the space of words, i.e. the Internet reservoir. In
order to check whether a critical phenomenon is taking place, we
have defined a susceptibiity-like parameter associated to the
order parameter $P$, and have shown that this parameter reaches a
peaked maximum in the neighborhood of the transition, what is
typical of a second order phase transition. In a further work, we
will address different reservoir sizes, by using not the worldwide
engine (google.com) but specific engines (german, spanish,
french,italian,...) whose characteristic reservoir sizes are
smaller, in order to make a finite size analysis of the
transition. The reservoir sizes of the specific engines will be
estimated through set theory. Finally, these results should be contrasted
with those of a purely stochastic process, in order to verify if the presence of such abrupt phenomenon is the result of a random
phenomenon.\\
On the other hand, the connections with Ramsey theory should be
studied in depth in future work.

\section{Acknowledgments}
The authors would like to acknowledge generous support from the National Science
Foundation, the Santa Fe Institute and grants FIS2009-13690 and S2009ESP-1691.

\end{document}